\documentclass[runningheads,hidelinks]{llncs}
\usepackage[utf8]{inputenc}
\usepackage{xspace}
\usepackage{booktabs}
\usepackage{hyperref}
\usepackage{amstext}
\usepackage{listings}
\usepackage{bold-extra}
\usepackage{booktabs}

\usepackage{float}
\newfloat{lstfloat}{htbp}{lop}
\floatname{lstfloat}{Listing}

\usepackage{tikz}
\usetikzlibrary{calc,shapes,fit,decorations.pathmorphing,decorations.pathreplacing,patterns,arrows,shapes}
\pgfdeclarelayer{background}
\pgfdeclarelayer{foreground}
\pgfsetlayers{background,main,foreground}
\tikzset{every edge/.style = {-stealth, draw}}
\tikzset{thick,>=stealth, shorten >=1pt}

\usepackage[inline]{enumitem}
\setlist{noitemsep}

\colorlet{rmcolor}{red!20}
\colorlet{kscolor}{green!20}
\colorlet{dscolor}{yellow!40}
\colorlet{awcolor}{black!20}

\newcommand{\RCE}{RCE\xspace}

\title{Automated and manual testing as part of the research software development process of \RCE}
\author{Robert Mischke\orcidID{0000-0003-3419-4897} \and Kathrin Schaffert\orcidID{0000-0002-9729-7730} \and Dominik Schneider\orcidID{0000-0003-4921-0693} \and Alexander Weinert\orcidID{0000-0001-8143-246X}\thanks{Authors are listed in alphabetical order}}
\institute{
Institute for Software Technology\\
German Aerospace Center (DLR) \\
51147 Cologne, Germany\\
\email{\{firstname\}.\{lastname\}@dlr.de}
}
\authorrunning{R. Mischke, K. Schaffert, D. Schneider, et al.}

\begin{document}

\maketitle

\begin{abstract}
	Research software is often developed by individual researchers or small teams in parallel to their research work.
  The more people and research projects rely on the software in question, the more important it is that software updates implement new features correctly and do not introduce regressions.
  Thus, developers of research software must balance their limited resources between implementing new features and thoroughly testing any code changes.

  We present the processes we use for developing the distributed integration framework RCE at DLR.
  These processes aim to strike a balance between automation and manual testing, reducing the testing overhead while addressing issues as early as possible.
  We furthermore briefly describe how these testing processes integrate with the surrounding processes for development and releasing.
\end{abstract}

\section{Introduction}
\label{sec:introduction}

More and more research is supported by software~\cite{BrettCroucherHainesEtAl2017}, which ranges wildly in scope and maturity.
Software may be developed and supported by major companies, it may result from internal research projects, or it may be a proof-of-concept script developed by individual researchers.
Particularly software resulting from internal research projects may be used by numerous research projects, while still being maintained by a handful of original developers.

To be able to implement new features required by new research projects, while simultaneously maintaining existing features, developers require effective and efficient software testing.
Such testing enables developers to detect issues early in development, thus simplifying their resolution.
Such testing, however, requires infrastructure and non-trivial resource investment to effectively spot issues~\cite{RTIHealthResearch2002}.
Developers should, e.g., not test features they implemented themselves, since they are likely to subconsciously avoid addressing edge cases~\cite{OliveiraRosenthalMorinEtAl2014}.

Thus, effective testing requires additional resources, which are unavailable in typical research projects.
Hence, these projects have to divide their existing resources between development of new features, maintenance of existing ones, and rigorous testing.
Each of these activities has a non-negligible overhead.

One such research software project is \RCE~\cite{BodenFlinkFoerstEtAl2021}, a distributed scientific integration framework which we develop at the Institute for Software Technology at the German Aerospace Center (DLR).
We describe the testing processes we employ to validate changes to \RCE and to recognize regressions as soon as possible.
These tests comprise automated tests as well as manual ones that address hard-to-test areas of \RCE's functionality.
We show how we balance the need to thoroughly test changes to \RCE with the overhead incurred by such tests.

\paragraph*{Related Work}
Software testing is a core topic discussed in virtually all education on software development~\cite{AmmannOffutt2016} as well as an active field of research~\cite{OrsoRothermel2014}.
Current research directions range from automated test case generation~\cite{AnandBurkeChenEtAl2013} over automated exploration of edge cases~\cite{BoehmePhamNguyenEtAl2017,LiZhaoZhang2018} to automated formal verification of software~\cite{Beyer2021}.
Most work in this area focuses on technical aspects of testing and verification, but does not consider embedding testing into software development processes.

The role of software engineers developing research software has been the topic of investigation in recent years~\cite{BrettCroucherHainesEtAl2017}.
To the best of our knowledge, there does not exist literature on testing such software effectively and efficiently.

Afzal, Alone, Glocksien, et al.\, have identified Software Test Process Improvement (STPI) approaches~\cite{AfzalAloneGlocksienEtAl2016} and studied selected STPI approaches at a large Swedish car manufacturer.
However, the STPI approaches that are effective and efficient for a large commercial company are likely very different from ones that are effective and efficient for research software engineering teams.

\paragraph*{Structure of this Work}
First, in Section~\ref{sec:background}, we give some background on \RCE and its userbase.
Subsequently, in Section~\ref{sec:preparation} we briefly describe the development process of \RCE and explain where testing takes place.
The focus of this work lies on Section~\ref{sec:testing}, where we present the technologies and processes we use for testing \RCE.
Afterwards, in Section~\ref{sec:releasing}, we briefly describe the process of releasing the tested changes to users before closing in Section~\ref{sec:conclusion} with a conclusion and an outlook on possible avenues for future work.

\section{Background}
\label{sec:background}

\RCE is a general tool for engineers and scientists from a wide range of disciplines to create and execute distributed workflows as well as evaluate the obtained results.~\cite{BodenFlinkFoerstEtAl2021}
It is available free of charge as an open source software at \url{https://rcenvironment.de/}, its source code is available at \url{https://github.com/rcenvironment}.
\RCE supports scientists and engineers to simulate, analyze and evaluate complex multidisciplinary systems like airplanes or satellites.
\RCE achieves this goal by enabling them to create automated and distributed workflows containing their own specific tools.
Thus, \RCE mainly serves to
\begin{enumerate*}[label=\arabic*)]
	\item integrate tools and compose them into workflows,
	\item share the integrated tools within a network, and to
	\item execute the developed workflows and manage the data flow across different network topologies.
\end{enumerate*}

A comprehensive introduction of \RCE and its features is out of scope of this work.
Instead, we refer to Flink, Mischke, Schaffert, et al.~\cite{FlinkMischkeSchaffertEtAl2022} for a user's description of \RCE and to Boden, Flink, Först, et al.~\cite{BodenFlinkFoerstEtAl2021} for technical details.

\begin{figure}
  \centering
  \includegraphics[width=\textwidth]{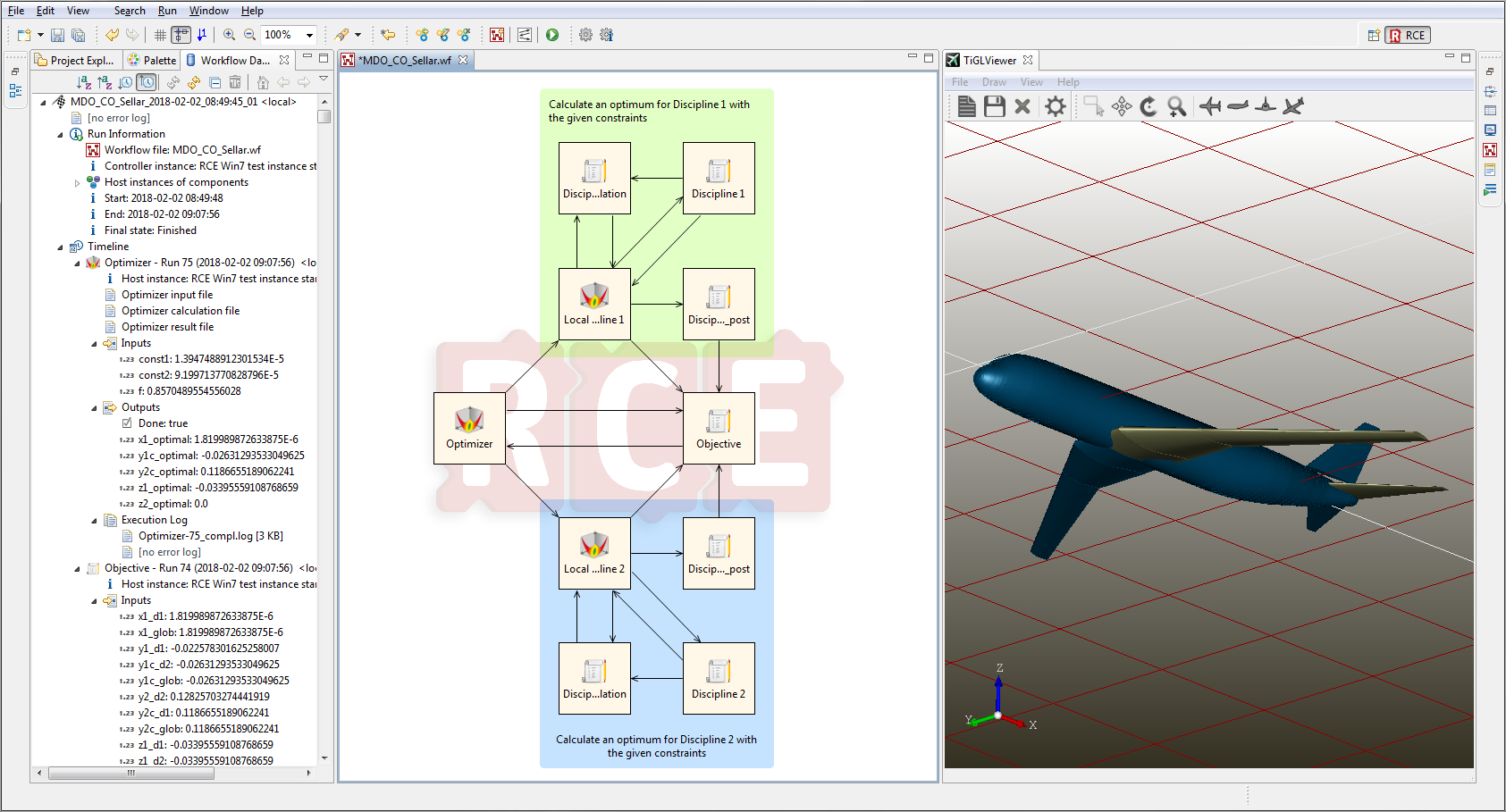}
  \caption{A screenshot of \RCE's GUI.}
  \label{fig:rce_gui}
\end{figure}

To implement the features described above in various IT environments, \RCE can be used \textit{interactively} via Graphical User Interface (GUI) or on servers as a \textit{headless} terminal-only application.
The interactive mode is typically used to design and execute workflows and to inspect their results.
The GUI is split up into \emph{views}, each supporting a different task requested by the user (cf. Figure~\ref{fig:rce_gui}).
The headless mode is typically used to provide tools in form of a tool server, or to forward communication between \RCE instances.
\RCE runs on Microsoft Windows and Linux operating systems in both modes.
Supporting multiple operating systems and window managers increases development and testing efforts.

Users usually work with an \RCE network consisting of multiple \RCE instances.
They integrate discipline-specific software tools into \RCE by defining their inputs and outputs and compose them into a workflow by matching types of inputs and outputs.
Thus, the outputs of one tool are used as inputs for another tool.
Tools can be shared in the network, whose topology is freely configurable.

One complication in testing \RCE arises from its diverse userbase, in terms of the activities and tasks~\cite{Norman2005} that users perform with \RCE as well as in terms of their prior knowledge.
A comprehensive classification of users with respect to these properties is out of the scope of this work.
We instead refer to Boden, Mischke, Weinert, et al.~\cite{BodenMischkeWeinertEtAl2020} for a more comprehensive classification.
Instead of focusing on the tasks and activities of users, we give a brief overview over the three most relevant types of users.
Furthermore, Boden, Flink, F{\"o}rst, et al. give an overview over research projects \RCE is involved in~\cite{BodenFlinkFoerstEtAl2021}.

Firstly, a part of \RCE's userbase consists of individual researchers from numerous fields who use \RCE on their local working machine for their respective research.
These users have diverse operating systems and system environments and typically interact with the GUI of \RCE.
Secondly, there exist groups of users comprised of researchers at different institutions that collaborate on a research project.~\cite{Risse2022}
These groups have built expansive \RCE-based server infrastructure connecting their respective organizational networks.
Finally, there exist research projects that use \RCE merely as an execution backend.
Researchers interact with \RCE via custom-made domain-specific interfaces~\cite{RisuenoBussemakerCiampaEtAl2020}.

\section{Development Process}
\label{sec:preparation}

Currently, the core development team of \RCE consists of four applied researchers, equivalent to about three full-time positions related to \RCE.
These developers are not only involved in the software development of \RCE, but are also embedded in research projects.
Here, they support and discuss current applications and use cases with users. 
They also collect feedback and input for roadmap planning.
Most members are also involved in non-development related tasks, e.g., they give software development training or supervise students.
Each of these activities has a non-negligible time commitment.

To avoid additional process overhead, we do not follow a prescribed development method, e.g., waterfall method, V-Model, or Scrum.
Instead, we put large emphasis on mutual communication, following the core ideas of agile software development~\cite{BeckBeedleVanBennekumEtAl2001}.
We coordinate collaboration in regular group and point-to-point meetings.
Within these meetings we continuously adapt the development roadmap to new or changing requirements using the \emph{Mantis} bug tracker~\cite{Mantis2022} for documentation and communication.
Further discussions are held as needed, e.g., for pair programming, knowledge transfer, or architectural decisions.

From theses meetings, the schedule for releasing new versions is generated and continuously adjusted.
There are dedicated meetings to decide about the scope of each release regarding new features, bugfixes, and/or other improvements.

Following semantic versioning~\cite{PrestonWerner2022}, releases fall into one of three categories:
\emph{Maintenance} releases only address issues with existing functionalities or contain internal changes.
\emph{Minor} releases may contain new features or changes to the user experience provided they do not break backwards compatibility.
Finally, \emph{major} releases may contain changes that potentially break backwards compatibility.

Before a release, we review all code changes made since the last release.
Additionally, we identify features that may have inadvertently been affected by these changes.
All new and modified parts of the code have to be tested carefully.
This can be done with automated and manual tests.
We describe the whole \RCE testing process in more detail in the following section.

\section{Testing}
\label{sec:testing}

Maintaining the quality of a complex software project like \RCE would be infeasible without automated tests.
Specifying and setting up these tests requires significant up-front effort.
However, once implemented, all automated tests can be applied at any time with minimal effort.
This addresses two important goals:

Firstly, any behavior of the software that is tested automatically does not need to be tested manually anymore.
Over the lifetime of a project, this saves large amounts of manual testing time.
This quickly offsets the initial setup effort.

Secondly, every existing test explicitly defines expectations about the system behavior.
An automated test setup allows these to be frequently validated.
This greatly reduces the risk of unintended changes in the system's behavior, e.g., due to code or environment changes.
These automated tests thus allow developers to change code without fear of breaking existing functionality.
Especially in a complex software with many dependencies like \RCE, this is crucial for project maintainability.
We describe our setup for automated tests in Section~\ref{sec:testing:automated}.

However, automated tests alone do not suffice to test \RCE comprehensively.
The more user-facing a feature is the more its expected behavior comprises simplicity, intuitiveness, and efficiency of its graphical representation.
These properties are subjective and thus hard to define in automated tests.
Furthermore, each automated test would have to be maintained when evolving the overall GUI.

Moreover, the GUI inherently depends on its system environment and thus, automated GUI-tests require sophisticated infrastructure.
In our experience, the cost of this infrastructure does not sufficiently benefit the bug-finding effort.

Finally, automated tests do not account for blind spots of the developer, e.g., the interaction of distinct GUI features or the overall consistency of the GUI.

To validate these requirements effectively and efficiently, we use manual test cases.
We write these in plain English and prompt the tester to explore existing behavior for regressions.
We describe our manual testing process in Section~\ref{sec:testing:manual}.

\subsection{Automated Tests}
\label{sec:testing:automated}

Automated tests can differ vastly regarding their implementation, test scope, and execution environment.
The automated tests used in \RCE team can be roughly divided into \emph{unit} tests, \emph{integration} tests, and \emph{behavior-driven} tests.

Unit tests validate expectations about isolated, fine-granular parts of the software's code.
Integration tests work on the combination of several code parts and usually test more complex interactions between them.
While this distinction seems clear in theory, it is often less clear-cut in practice.
Thus, we combine both types in our code repository with no explicit technical distinction between them.
So far, we have not encountered any downsides from this approach.

More relevant, in contrast, is the distinction between fast-running and long-running tests.
Combining both would slow down the execution of the fast-running tests, and lead to them being executed less often.
To avoid this, we manually mark long-running test cases on the code level.
Our testing infrastructure is configured to exclude these from the frequent standard test runs, which are executed multiple times daily whenever there are code changes.
The slow tests, instead, are only executed at longer intervals, e.g., about weekly.

Both types of tests use the \emph{JUnit} framework~\cite{JUnit2022}, with integration tests additionally using \emph{EasyMock}~\cite{EasyMock2022} for setting up more complex test scenarios.

Unit and integration tests are used by many software projects, and can be considered a standard practice in today's software development.
Even integration tests, however, only work on relatively small subsets of the overall application.
For this reason, we complement them with behavior-driven tests to validate high-level, functional aspects of \RCE.
Such test cases include the verification of
\begin{itemize}
	\item the execution and outcome of test workflows, including both success and intentional failure cases,
	\item the proper startup process of the application itself, e.g., for detecting stability or concurrency issues,
	\item the effect of certain command line parameters on the application,
	\item the resulting state after connecting \RCE instances via their network features,
	\item the effect of authorization operations, e.g., changing access control settings of workflow components,
	\item and the presence or absence of specific log entries and/or warnings.
\end{itemize}

Listing~\ref{lst:bdd-basic} shows a very simple example of such a behavior-driven test case.
It is written using the \emph{Gherkin} syntax~\cite{Gherkin2022}, which is a standardized text format for specifying human-readable test cases.
A core aspect of this syntax is the standardization of test flows by breaking them down into ``Given/When/Then'' clauses.
These clauses define the test environment or preconditions, test activities to be performed, and expected outcomes, respectively.
``And'' clauses are a syntactical addition for better readability, and repeat the type of the previous clause. Lines starting with ``@'' are tags that can be used for test organization, e.g., for identifying individual tests or defining sets of tests to execute together.

\begin{lstfloat}
\begin{lstlisting}[
	label={lst:bdd-basic},
	caption={A basic behavior-driven test case for \RCE. The symbol \mbox{\textcolor{gray}{$\hookrightarrow$}} does not occur in the actual code but indicates an added linebreak for the sake of presentation.},
	captionpos=b,
	basicstyle=\ttfamily\scriptsize,
	frame=single,
	morekeywords={Scenario, Given, When, Then, And},
  breaklines=true,
  breakatwhitespace=true,
  postbreak=\mbox{\textcolor{gray}{$\hookrightarrow$}\space},
  breakindent=1.5cm
]
@Network02
@NoGUITestSuite
Scenario: Basic networking between three instances (auto-start connections, no relay flag)

Given instances "NodeA, NodeB, NodeC" using the default build
And   configured network connections "NodeA->NodeC [autoStart], NodeB->NodeC [autoStart]"

When  starting all instances
Then  all auto-start network connections should be ready within 20 seconds 
And   the visible network of "NodeA" should consist of "NodeA, NodeC"
And   the visible network of "NodeB" should consist of "NodeB, NodeC"
\end{lstlisting}
\end{lstfloat}

For executing tests written in this syntax, we use the Java version of the \emph{Cucumber} framework~\cite{Cucumber2022}.
As the activity and result phrases (e.g., ``starting all instances'' and ``should consist of'') are highly application-specific, these can not be provided by the generic testing framework.
Instead, it provides tools to define them, and to map them to specific execution code, which has to be written manually.
The advantage of this is that the relatively complex execution code is only created once.
Once this is done, the semantic behavior to be tested can be described and maintained with much lower effort.
Additionally, these semantic test descriptions are much easier to read, understand, and validate than equivalent tests defined in program code.
To a certain degree, this even allows test cases to be written and maintained by non-developers autonomously.

To ensure that the actual behavior of the complete application is tested, all behavior-driven tests are run on standalone product builds of the application code.
This required building a fairly complex infrastructure to download, configure, and start/stop application instances automatically.
Just like implementing the semantic phrases described above, however, this was a one-time effort.
While we cannot quantify this exactly, we are certain from observation that this was quickly offset by reduced test creation time and lower manual testing effort.

\subsection{Manual Tests}
\label{sec:testing:manual}

In this section we describe our manual testing process that we have established over the last years.
This process has proven successful for our distributed and small developer team.
Moreover, working from home became more popular during the COVID-19 pandemic, which further increased the distribution of development teams.
First, we will describe the roles we have identified in our project, then we will give a brief overview of our general testing process.
After that, we will have a deeper look into the process of a test session. 
Finally, we will present the main principles for such sessions that we found for the \RCE project.

\subsubsection{Roles in Manual Testing}
\label{sec:testing:manual:roles}

In this section we introduce the roles involved in our manual testing process, namely the \emph{developers}, the \emph{testers} and the \emph{test manager}.
	
In our experience it is unusual for research software projects to have a dedicated testing team.
The typically small team size along with the motivation to produce high-quality research results means that teams concentrate on conducting their research tasks.
Often, software testing---particularly manual testing---is not given the required attention.
Nevertheless, manual tests are inevitable for developing high-quality software due to the reasons described above.

Below we describe our approach to manage manual test phases in our small team with high workload.
When we talk about small teams, we mean teams with fewer than ten team members.
In previous years, the \RCE team was about the size of four to ten team members.
Due to our team size, we do not usually split into developers and testers, i.e., every developer is also a tester and vice versa.
At least during the release phases when manual testing becomes more important, these roles overlap.
The dedicated role of the test manager coordinates the manual testing process.
In \RCE, the test manager is also part of the development team and thus even takes on three roles during testing.

The developers are responsible for providing test cases for their work.
This has two aspects.
On the one hand, they write automated tests as described in Section~\ref{sec:testing:automated}.
On the other hand, they identify software parts for manual testing, write manual test cases, and inform the test manager about them.

The test manager reviews the tests and provides feedback.
In addition, the test manager schedules and prepares the test sessions and guides the team through them.
They also track the test progress and keep an eye on any unfinished tasks for both developers and testers.
To manage the overall testing process, we use the test management software \emph{Testrail}~\cite{Testrail2022}. 
Testrail is a web-based management solution that lets users manage, track and organize test cases and test plans.
During test sessions, the testers are responsible for completing the test cases assigned to them by the test manager.
Furthermore, they give feedback on the test results to the team, especially to the responsible developer. 
Since every tester is also a developer, the team is doubly challenged during such phases.
First, everyone needs to test the application, while second, development activities---such as bug fixes or further improvements---need to continue.
It is a special challenge to organize the small team and the increased workload well during this time.
Especially when the release has to meet a certain deadline, it is critical to avoid personal bottlenecks as much as possible, which can be hard as not all tasks can be easily shifted between team members.

\subsubsection{Manual Testing Process}
\label{sec:testing:manual:process}

To minimize the time for manual testing, we test in several phases: The \emph{Pretesting}, the \emph{Release Testing} and the \emph{Final Testing}.

Each new feature must pass a Pretesting phase.
Depending on the feature size this pretest takes a few hours to one or two days.
One or two testers are recruited from the development team to test the new feature on a maximum of two operating systems.
These pretests are used to detect and correct as many errors as possible before the actual testing phase.
Only when these tests are passed the new feature is discussed for inclusion in the next release.
This procedure reduces the test effort during the Release Testing to a minimum.
Minor developments, e.g., bug fixes or smaller improvements, are only pretested by the developer.

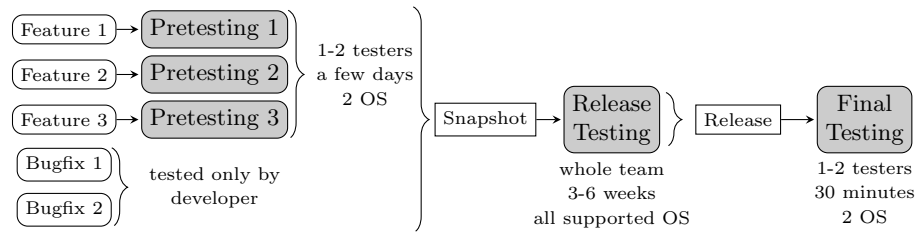
\begin{figure}
  \centering
  \begin{tikzpicture}[xscale=0.8,yscale=.6]
    
	\node(feature1)[draw,rounded corners] at (0,0) {\scriptsize Feature 1};
	\node (pretesting1)[draw,fill=black!20,rounded corners] at (2.5,0) {\footnotesize Pretesting 1};
	\node(feature2)[draw,rounded corners] at (0,-1) {\scriptsize Feature 2};
	\node(pretesting2)[draw,fill=black!20,rounded corners] at (2.5,-1) {\footnotesize Pretesting 2};
  \node(feature3)[draw,rounded corners] at (0,-2) {\scriptsize Feature 3};
	\node(pretesting3)[draw,fill=black!20,rounded corners] at (2.5,-2) {\footnotesize Pretesting 3};
	\node(text1)[align=center,execute at begin node=\setlength{\baselineskip}{1em}] at (5,-1) {\scriptsize 1-2 testers\\\scriptsize a few days\\\scriptsize 2 OS};
	\node(bug1)[draw,rounded corners] at (0,-3) {\scriptsize Bugfix 1};
	\node(bug2)[draw,rounded corners] at (0,-4) {\scriptsize{Bugfix} 2};
	\node(text2)[align=center,execute at begin node=\setlength{\baselineskip}{1em}] at (2.5,-3.5) {\scriptsize tested only by\\ \scriptsize developer};
	\node(snap)[draw] at (7,-2) {\scriptsize Snapshot};
	\node(release Test)[draw,fill=black!20,rounded corners,align=center] at (9.1,-2) {\footnotesize Release \\ Testing};
	\node(text3)[align=center,anchor=north,execute at begin node=\setlength{\baselineskip}{1em}] at (release Test.south) {\scriptsize whole team\\\scriptsize 3-6 weeks\\ \scriptsize all supported OS};
	\node(release)[draw] at (11.2,-2) {\scriptsize Release};
	\node(final Test)[draw,rounded corners,fill=black!20,align=center] at (13.3,-2) {\footnotesize Final \\ Testing};
	\node(text4)[align=center,anchor=north,execute at begin node=\setlength{\baselineskip}{1em}] at (final Test.south) {\scriptsize 1-2 testers\\\scriptsize 30 minutes\\\scriptsize 2 OS};
	
	\draw[->] (feature1) -- (pretesting1);
	\draw[->] (feature2) -- (pretesting2);
	\draw[->] (feature3) -- (pretesting3);
	\draw[->] (snap) -- (release Test);
	\draw[->] (release) -- (final Test);
	
	\draw[decorate,decoration={brace,amplitude=6pt}] (pretesting1.north east) -- (pretesting3.south east);
	\draw[decorate,decoration={brace,amplitude=6pt}] (bug1.north east) -- (bug2.south east);
	\draw[decorate,decoration={brace,amplitude=6pt}] (5.8,0.5) -- (5.8,-4.5);
	\draw[decorate,decoration={brace,amplitude=6pt}] (10,-1.4) -- (10,-2.6);
\end{tikzpicture}
  \caption{The manual testing process. Nodes with gray background denote testing activities, while rounded nodes with white background denote code changes. Square nodes with white background denote software artifacts.}
  \label{fig:testprocess}
\end{figure}

The main testing phase is the Release Testing, where all new developments come together and are tested in their entirety.
The new developments comprise not only new features, but also improvements to existing features and bug fixes.
Each Release Testing lasts several weeks and ends when the team decides that the software under test is ready for release.
We describe how the team comes to this decision in the next section.
The duration of the Release Testing does typically not exceed four weeks.
In contrast to Pretestings, the entire team usually participates in these tests.
We test up to ten configurations, each comprising an operating system, a desktop environment, a certain Java Runtime (JRE) version, and the choice whether to execute \RCE with its GUI or in headless mode.

At the end of the Release Testing the final product is built.
Before it is officially released, we apply the Final Testing to guard against the possibility of otherwise undetected errors having crept in.
There could, e.g., have been upgrades of build tools or system libraries, IT problems like running out of disk space causing build steps to finish incompletely, or simply code changes causing unexpected side effects which have not been caught by test cases.
To have a chance of detecting these, two team members take one last look at the product.
This takes about half an hour and focuses on the basic functionalities of \RCE.

We now take a closer look at the Release Testing phases.

\subsubsection{Test Session Process} 
\label{sec:testing:session:process}

Before Release Testing there are some organizational tasks for the test manager.
First, they check whether manual test cases exist for all new features and minor enhancements included in the release.
If so, the test plan can be created.
Otherwise, the test manager reminds the responsible developers to add their test cases.
In addition, the test manager discusses the assignment of testers to platforms so that the testers can prepare their testing environment.
As we test for different operating systems, this typically involves setting up one or more virtual machines to simulate these platforms.

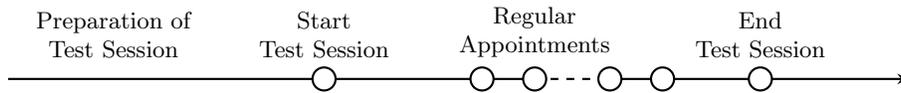
\begin{figure}
  \centering
  \begin{tikzpicture}[thick]

  \path[draw,-] (0,0) --
    node[circle,pos=.2,label={[align=center]above:{Preparation of \\ Test Session}}] {}
    node[draw,fill=white,circle,pos=.6,label={[align=center]above:{Start\\Test Session}}] {}
    node[draw,fill=white,circle,pos=.9] {}
    node[draw,fill=white,circle,pos=1.0,label={[align=center]above:{Regular\\Appointments}}] {}
    (7,0);
  \begin{pgfonlayer}{background}
    \path[draw,-,dashed,thick] (7,0) -- (8,0);
  \end{pgfonlayer}
  \path[draw,->,>=stealth,thick] (8,0) --
    node[draw,fill=white,circle,pos=0] {}
    node[draw,fill=white,circle,pos=0.175] {}
    node[draw,fill=white,circle,pos=0.5,label={[align=center]above:{End\\Test Session}}] {}
    (12,0);

\end{tikzpicture}
  \caption{Test Session Process}
  \label{fig:testsession}
\end{figure}

The testing then starts with a kick-off meeting.
In this meeting, the testers discuss all necessary details for the testing as a team and clarify any questions, e.g., final organization issues or comprehension questions for specific test cases.
Developers share knowledge about their new features, point out what aspect testers should focus on, or give a brief overview of a new functionality.
The test manager moderates the meetings and reminds the team of our common principles for testing.
We discuss these principles in detail in the following section.

During the test session the whole teams meets twice a week for about half an hour each.
These meetings serve to focus the team on the most important tasks for the days ahead.
The team takes a look at the test progress together.
The test manager presents intermediate test results and checks for ``blind spots'' that have not yet been tested sufficiently.
Testers draw the team's attention to critical bugs they have detected, especially if those bugs prevent further testing.
Developers report on bug fixes and point out test cases to be retested.

Beyond that, we use these meetings to prevent employee overload.
This is especially important for small teams, where each team member performs multiple functions.
If someone in the team has so much development work that the test scope can no longer be met, the test manager coordinates support.
The test manager adjusts the test plan accordingly and reassigns the test cases.

As the test phase progresses, fewer and fewer tasks are outstanding.
This phase has no pre-determined deadline, but rather comprises iteratively resolving and retesting critical bugs.
Towards the end, everyone ensures no important tasks have been overlooked.
Once Release Testing has ended, the team cleans up, at most two team members perform Final Testing, and we begin review.

During this review, the entire team gathers feedback, suggestions and ideas on how we can optimize our process in the future.
The test manager then evaluates this collection and develops proposals that are presented to the team.
This review process continuously contributes to the improvement of our testing process.

\subsubsection{Principles for Testing} 
\label{sec:testing:manual:principles}

Within the \RCE project we have agreed on some testing principles.
These are to
\begin{enumerate*}[label=\arabic*)]
	\item\label{prin:final-state} ensure that assigned test cases reach a \emph{final state}, to
	\item\label{prin:life-cycle} stick to the \emph{Test Case Life Cycle}, and to 
	\item\label{prin:box} look outside the box.
\end{enumerate*}

Testers have to take care that all their assigned test cases have been executed at the end of the test session (Principle~\ref{prin:final-state}. 
During the test session a test case can reach different states. 
We list all possible states and their meanings in Table~\ref{tab:test-case-states}.

\begin{table}
  \caption{States of test cases.}
  \label{tab:test-case-states}
  \scriptsize
  \begin{tabular}{lp{.7\textwidth}} \toprule
    \textbf{Passed} & Test case was executed and no errors were observed \\
    \textbf{Passed with Remarks} & Similar to ``Passed'', but the tester discovered some improvements for a future release \\
    \textbf{Not applicable} & Test case was not executed because it is not applicable on the current test configuration \\
    \textbf{Won't test} & Test case was not executed because it has been tested sufficiently on other configurations  \\
    \textbf{Failed} & Test case failed \\
    \textbf{Failed \&  Blocked} & Test case failed and blocked on all other configurations \\
    \textbf{Retest} & Test case ready to retest \\
    \textbf{Waiting for new build} & Test case ready to retest in the next build \\
    \textbf{Blocked} & Test case blocked by the developer due to ongoing development work \\
    \textbf{Failed \& Postponed} & Test case failed and fix is postponed to a future release \\ \bottomrule
  \end{tabular}
\end{table}

\begin{figure}
  \centering
  \includegraphics[width=12cm]{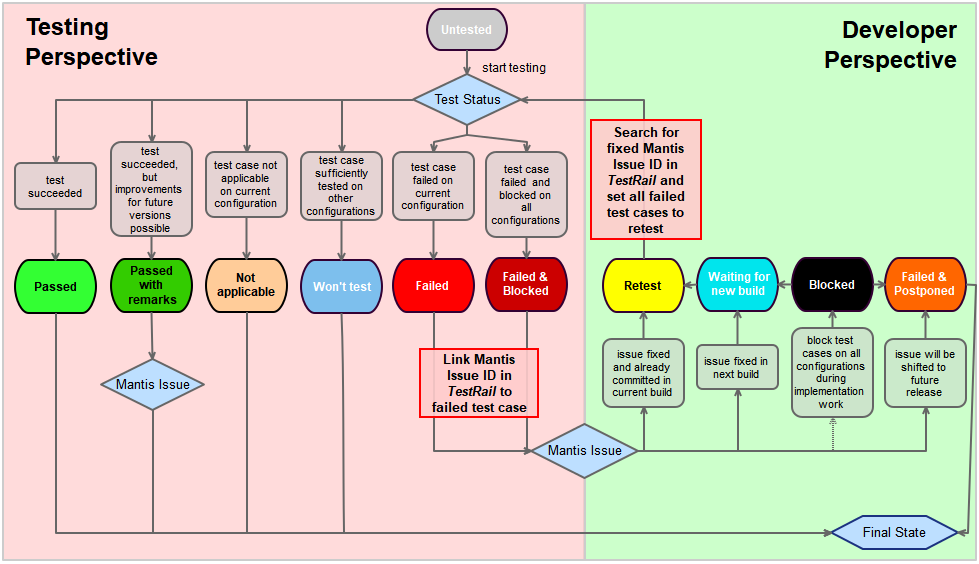}
  \caption{The Test Case Life Cycle}
  \label{fig:lifecycle}
\end{figure}

States are final states if they cannot reach any other state afterwards.
Final states are for example ``passed'', ``won’t test'' or ``failed \& postponed''.
When a test session starts, all test cases are ``untested''.
Tester cycles them through different states before they reach a final state. 
In doing so they follow the Test Case Life Cycle (Principle~\ref{prin:life-cycle}.
Some states lead to an action that must be performed by the tester or developer while others are intermediate or final states.
Figure~\ref{fig:lifecycle} shows all states with their actions.
The states on the left-hand side can be assigned by the testers.
The states on the right-hand side are to be set exclusively by the responsible developers or by the test manager.

We now give an example of a possible life cycle of a test case. 
A tester picks an ``untested'' test case and executes it. 
During testing the tester finds a bug in the tested feature, sets the test case to ``failed'' and creates a Mantis issue.
The tester references the Mantis issue ID in the failed test case. 
This issue describes the erroneous behavior in detail and is assigned to the responsible developer.
During the regular appointments, the tester reminds the team about the found bug if it has not been noticed yet. 
Here the team also discusses the priority of found bugs and whether they need to be fixed for the current release.
If this is not the case, the test case is set to ``failed \& postponed'', a final state. 
Otherwise the responsible developer fixes the bug and sets the status to ``waiting for new build''. 
The associated test case can be easily found via the issue reference. 
With the next build, the developer or the test manager change the state of the test case to ``retest''. 
This status, in turn, triggers the tester to perform the test once again. 
This cycle can be repeated several times until the test is finally ``passed''.

The states presented here were established iteratively via our review process. 
Using of these states has proven successful in the development of \RCE.

Finally, we have the commitment in the team that test cases should not be understood as a step-by-step instruction.
Rather, they indicate an area or functionality of the software to be tested and provide suggestions on how to test. 
Often, of course, there are some instructions to follow, e.g., to create the required test setup. 
Apart from that, every tester should always think outside the box (Principle~\ref{prin:box} and also test functionalities ``close'' to the described one. 
They should keep in mind the following questions:
\begin{enumerate*}[label=\arabic*)]
	\item Is there anything I could test that is not described in the test case?
	\item Are there interactions with other functionalities not considered in the tests?
	\item How can I try to abuse the functionality?
\end{enumerate*}
This contributes to exploratory testing and increases test coverage.

\section{Releasing}
\label{sec:releasing}

In the previous section we have described how we validate the changes included in a new version of \RCE.
After having done so, we then move to release this new version to users.
This requires three steps:
First, we build the final artifacts that are distributed to the users.
Afterwards, we deploy these artifacts to the download site.
In the last step we inform existing and potential users about this new version.
In this section, we give a brief overview over each of these steps.

First, we construct the artifacts to be distributed using \emph{Jenkins}~\cite{Jenkins2022} and \emph{Maven}~\cite{Maven2022}.
This pipeline compiles the codebase into
\begin{enumerate*}[label=\arabic*)]
  \item an update site for automatic updates to existing \RCE installations,
  \item zip files for both Windows and Linux, both digitally signed,
  \item a .deb package as well as a Debian repository for installation on Debian-based Linux distributions, and
  \item an .rpm package for installation on Red Hat-based Linux distributions.
\end{enumerate*}

We publish these artifacts to our website at \url{https://rcenvironment.de}.
We moreover mirror the source code of the newly released artifact to Github at \url{https://github.com/rcenvironment} for accessibility and create a Github release containing this artifact.

Finally, we inform users about the newly released version.
To do so, we first collect a list of changes with respect to the previous version.
For this, we consult both the Mantis changelog as well as a manually compiled list of new features created by developers in the wiki.
We inform users by compiling and publishing
\begin{enumerate*}[label=\arabic*)]
  \item a technical changelog on Github,
  \item a nontechnical changelog on rcenvironment.de,
  \item an email newsletter pointing users to the nontechnical changelog, and
  \item a tweet at \url{https://twitter.com/RCEnvironment} pointing users to the nontechnical changelog.
\end{enumerate*}
We publish these items in order, where each item contains a reference to one or more items published before.

\section{Conclusion and Future Work}
\label{sec:conclusion}

In this work we have presented our processes for ensuring that changes made to \RCE implement the desired feature, integrate well with ``neighboring'' features, and do not cause regressions in already existing features.
We have shown how we test changes, bugfixes, and new features affecting various aspects of \RCE, from its technical foundations to user-facing features like its GUI.

In future work, we are looking to expand the scope of automated behavior-driven testing of \RCE.
In particular, we aim to improve the scalability of our distributed test scenarios.
To this end, our test orchestration setup shall be adapted to make use of cloud infrastructure.
This will allow larger test scenarios to be defined and executed, and create more options for intense load testing.  

Moving to this cloud-based infrastructure also opens up possibilities for new testing approaches.
First steps have already been taken towards employing aspects of \emph{Chaos Testing}, e.g., actively introducing network and/or instance failures into test networks to verify \RCE's robustness against them.~\cite{Hampel2021}
Thus, we will be able to replicate real-world circumstances more closely, which will allow even more thorough testing, leading to increased robustness of the final product.

We are moreover looking to introduce further automation for testing \RCE's GUI.
Automated GUI testing is notorious for being difficult to get right, as there are several conflicting aspects to balance.
On one hand, such tests must be sensitive enough to actually detect functional regressions.
On the other hand, they should be flexible enough to accept minor changes to the GUI that affect neither the functionality nor the aesthetics of the GUI significantly.
Moreover, the resulting tests should be maintainable enough to be easily adapted to new features or desired changes to the GUI.
Nonetheless, given \RCE's large and fairly complex GUI, we could benefit a lot from automating parts of its testing.
Therefore, based on earlier attempts, we plan to reevaluate the current state of testing Eclipse-based GUI applications.

\bibliographystyle{splncs04}
\bibliography{se4science2022-rce-releasing}

\end{document}